\documentclass [a4paper,fleqn]{article}
\usepackage{graphicx}
\usepackage {subfigure,epsfig}

\usepackage {amsmath} \usepackage{amssymb} \usepackage{cite} \usepackage{amsthm}
\numberwithin{equation}{section} \numberwithin{table}{section} \mathindent=0pt
\theoremstyle{plain} \newtheorem{theorem}{Theorem}

\numberwithin{theorem}{section}

\begin{document}

\title{\textbf{Special polynomials associated with the fourth order analogue
to the Painlev\'{e} equations}}

\author{Nikolai A. Kudryashov, Maria V. Demina }

\date{Department of Applied Mathematics\\
Moscow Engineering and Physics Institute\\ (State University)\\
31 Kashirskoe Shosse, 115409, Moscow, \\ Russian Federation}
\maketitle

\begin{abstract}
Rational solutions of the fourth order analogue to the Painlev\'{e}
equations are classified. Special polynomials associated with the
rational solutions are introduced. The structure of the polynomials
is found. Formulas for their coefficients and degrees are derived.
It is shown that special solutions of the Fordy - Gibbons, the
Caudrey - Dodd - Gibbon and the Kaup - Kupershmidt equations can be
expressed through solutions of the equation studied.
\end{abstract}



\section{Introduction}

It is well known that the general solutions of the six Painlev\'{e}
equations $(P_1-P_6)$ can not be expressed through known elementary
or classical special functions in the general case because their
solutions determine new transcendental functions. However, the
equations $P_2-P_6$ possess hierarchies of rational and algebraic
solutions at certain values of the parameters.

A. I. Yablonskii and A. P. Vorob'ev were first who expressed the
rational solutions of $P_2$ via the logarithmic derivative of the
polynomials, which now go under the name of the Yablonskii --
Vorob'ev polynomials \cite{Yablonskii01, Vorob'ev01}. Later K.
Okamoto suggested special polynomials for certain rational solutions
of $P_4$ \cite{Okamoto01}. H. Umemura derived analogues polynomials
for some rational and algebraic solutions of $P_3$  and $P_5$
\cite{Umemura01}. All these polynomials possess a number of
interesting properties. For example, they can be expressed in terms
of Schur polynomials. Besides that the polynomials arise as the
tau-functions and satisfy recurrence relations of Toda type.
Recently these polynomials have been intensively studied
\cite{Clarkson01, Clarkson02, Clarkson03, Nuomi01, Kajiwara01,
Kaneko01}.

Not long ago P. A. Clarkson and E. L. Mansfield suggested special
polynomials for the equations of the $P_2$ hierarchy
\cite{Clarkson04}. Also they studied the location of their roots in
the complex plane and showed that the roots have a very regular
structure.

The aim of this work is to introduce special polynomials related to
rational solutions of the following equation, which is analogue to
the Painlev\'{e} equations

\begin{equation}
\label{1.1}
w_{zzzz}+5\,w_{z}\,w_{zz}-5\,w^2\,w_{zz}-5\,w\,w_{z}^{2}+w^5-z\,w-\beta=0.
\end{equation}

Originally this equation was found from the Fordy -- Gibbons
equation \cite{Fordy01}

\begin{equation}
\label{1.2}\omega_t+\omega_{xxxxx}+5\,\omega_{x}\,\omega_{xxx}-5\,\omega^2\,\omega_{xxx}
+5\,\omega_{xx}^{2}-20\,\omega\,\omega_x\,\omega_{xx}-5\,\omega_{x}^{3}+5\,
\omega^4\,\omega_x=0
\end{equation}
through the scaling reduction

\begin{equation}\begin{gathered}
\label{1.3}\omega(x,t)=(5\,t)^{-\frac15}\,w(z),\quad
z=x\,(5\,t)^{-\frac15}.
\end{gathered}\end{equation}

Equation \eqref{1.1} was first considered in \cite{Hone01} and later
in works \cite{Kudryashov01, Kudryashov02, Kudryashov03,
Kudryashov04, Kudryashov05, Mugan01, Cosgrove01}. This equation has
a number of properties similar to that of the Painlev\'{e}
equations. More exactly it possesses the B\"{a}cklund
transformations, the Lax pair, rational and special solutions at
certain values of the parameter $\beta$ \cite{Hone01, Kudryashov01}.
These special solutions are expressible in terms of the first
Painlev\'{e} transcendent \cite{Kudryashov02, Kudryashov05}.  The
Cauchy problem for this equation can be solved by the isomonodromic
deformation method.

Let us demonstrate that special solutions of the Caudrey -- Dodd --
Gibbon equation can be expressed through solutions of \eqref{1.1}.

The Caudrey -- Dodd -- Gibbon equation (or Savada - Kotera equation)
can be written as \cite{Weiss01,Caudrey01, Savada01}

\begin{equation}\begin{gathered}
\label{1.4}u_t+u_{xxxxx}-5\,u\,u_{{{xxx}}}-5\,u_x\,u_{{{xx}}}+5\,u^2\,u_x=0.
\end{gathered}\end{equation}

This equation has the self -- similar reduction
\begin{equation}\begin{gathered}
\label{1.5}u(x,t)=(5\,t)^{-\frac25}\,y(z),\quad
z=x\,(5\,t)^{-\frac15},
\end{gathered}\end{equation}
where $y(z)$ satisfies the equation
\begin{equation}\begin{gathered}
\label{1.6}y_{{{zzzzz}}}-5\,y\,y_{{{zzz}}}-5\,y_{{z}}y_{{{zz}}}+5\,{y}^{2}y_{{z}}-
2\,y-z\,y_{{z}}=0.
\end{gathered}\end{equation}
The Miura transformation $y(z)=w_{{z}}+{w}^{2}$ relates solutions of
\eqref{1.6} to solutions of the equation
\begin{equation}\begin{gathered}
\label{1.8}\left(\frac{d}{dz}+2w\right)\frac{d}{dz}(w_{{{
zzzz}}}+5\,w_{{z}}\,w_{{{zz}}}-5\,{w}^{2}\, w_{{{ zz}}}
-5\,w\,{w_{{z}}}^{2}+{w}^{5}-w\,z-\beta)=0.
\end{gathered}\end{equation}
Thus we see that for any solution of \eqref{1.1} there exists a
solution of \eqref{1.4}.

The Kaup - Kupershmidt equation \cite{Kaup01, Weiss01}
\begin{equation}\begin{gathered}
\label{1.9}v_t+v_{xxxxx}+10\,v\,v_{xxx}+25\,v_{x}\,v_{xx}+20\,v^2\,v_x=0
\end{gathered}\end{equation}
also possesses solutions, which can be expressed via solutions of
\eqref{1.1}. Indeed it has the self -- similar reduction
\begin{equation}\begin{gathered}
\label{1.10}v(x,t)=(5\,t)^{-\frac25}\,y(z),\quad
z=x\,(5\,t)^{-\frac15}
\end{gathered}\end{equation}
with $y(z)$ satisfying the equation
\begin{equation}\begin{gathered}
\label{1.11}y_{zzzzz}+10\,y\,y_{{{zzz}}}+25\,y_{{z}}y_{{{zz}}}
+20\,{y}^{2}y_{{z}}-z\,y_{{z}}-2\,y=0.
\end{gathered}\end{equation}
After making the Miura transformation $y(z)=w_{{z}}-\frac12{w}^{2}$
we obtain
\begin{equation}\begin{gathered}
\label{1.13}\left(\frac{d}{dz}-w\right)\frac{d}{dz}(w_{{{
zzzz}}}+5\,w_{{z}}\,w_{{{zz}}}-5{w}^{2}\,w_{{{ zz}}}
-5w\,{w_{{z}}}^{2}+{w}^{5}-w\,z-\beta)=0.
\end{gathered}\end{equation}

Hence the Fordy - Cibbons equation \eqref{1.2}, the Caudrey -- Dodd
-- Gibbon equation \eqref{1.4} and the Kaup - Kupershmidt equation
\eqref{1.9} admits solutions in terms of solutions of \eqref{1.1}.

Also we would like to mention that apparently the equation
\eqref{1.1} defines new transcendental functions like the
Painlev\'{e} equations do.

\section{Special polynomials associated with rational solutions of equation \eqref{1.1}}

Let us briefly review some facts concerning the equation
\eqref{1.1}, which we will need late. Let $w\equiv w(z;\beta)$ be a
solution of \eqref{1.1}. Then the transformations
\begin{equation}\begin{gathered}
\label{2.1}T_{2-\beta}:\,w(z;2-\beta)=w+\frac{2\beta-2}{z-w_{zzz}+ww_{zz}-
3w_{z}^{2}+4w^2w_z-w^4} \hfill\\
T_{-1-\beta}:\,w(z;-1-\beta)=w+\frac{2\beta+1}{z+2w_{zzz}+4ww_{zz}+3w_{z}^{2}-2w^2w_z-w^4}
\end{gathered}\end{equation}
generate other solutions of \eqref{1.1}, provided that $\beta\neq1$
for $T_{2-\beta}$ and $\beta\neq-1/2$ for $T_{-1-\beta}$
\cite{Kudryashov02}.

Let $z=z_0$ be a pole of the solution $w(z;\beta)$, then the Laurent
expansion of $w(z;\beta)$ in a neighborhood of $z_0$ is the
following
\begin{equation}\begin{gathered}
\label{2.2}w(z;\beta)=\frac{e}{z-z_0}+c^{(e)}(z-z_0)+\phi^{(e)}(z-z_0),
\end{gathered}\end{equation}
where $e$ takes one of the values $1$, $4$, $-2$, $-3$ and
$\phi^{(e)}(z-z_0)\sim o(z-z_0)$ is a holomorphic function in a
neighborhood of $z_0$. In fact, there are four types of Laurent
expansion around a movable pole.

If now the point $z=\infty$ is a holomorphic point of $w(z;\beta)$,
in particular it is the case of rational solutions, then the
expansion of $w(z;\beta)$ around infinity is
\begin{equation}
\label{2.3}w(z;\beta)=-\frac{\beta}{z}+\sum_{l=1}^{\infty}
c_{\beta,-5l-1}z^{-5l-1}.
\end{equation}
All the coefficients $c_{\beta,-5l-1}$ in \eqref{2.3} can be
sequently found. The first few of them are given below
\begin{equation}\begin{gathered}
\label{2.3a}c_{\beta,-6}=-{\beta\, \left( \beta+4 \right)  \left(
\beta-2 \right)  \left( \beta-3 \right)
 \left( \beta+1 \right) },\hfill\\
c_{\beta,-11}=(5\,{\beta}^{4}-295\,{\beta}^{2}+270\,\beta
+3024)c_{\beta,-6},\hfill\\
c_{\beta,-16}=(35\,{\beta}^{8}-6270\,{\beta}^{6}+6560\,{\beta}^{5
}+380055\,{\beta}^{4}-467400\,{\beta}^{3}\\
 -9286740\,{\beta}^{2} +
8845200\,\beta+72648576)c_{\beta,-6}.
 \end{gathered}
\end{equation}
What is more, the ratio $c_{\beta,-5l-1}/c_{\beta,-6}$ is a
polynomial in $\beta$ for all $l>1$. The expansions \eqref{2.2},
\eqref{2.3} can be obtained with the help of algorithms of power
geometry. For more information see \cite{Bruno01, Bruno02,
Kudryashov07}.

Rational solutions of the equation \eqref{1.1} are classified in the
following theorem.
\begin{theorem}
\label{T:2.1} The equation \eqref{1.1} possesses rational solutions
if and only if $\beta \in \mathbb{Z}/\{1\pm3k$, $k \in
\mathbb{N}\cup0\}$. They are unique and have the form
\begin{equation}
\begin{gathered}
\label{2.4}w(z;\beta_n^{(1)})=(-1)^{n}\frac{d}{dz}\ln
{\left(\frac{Q_{n-1}}{Q_n} \right)},\\
w(z;\beta_n^{(2)})=(-1)^{n-1}\frac{d}{dz}\ln {\left(\frac{R_{n-1}}
{R_n}\right)},
\end{gathered}
\end{equation}
where $Q_n(z)$ and $R_n(z)$ are polynomials, $n\in \mathbb{N}$  and
\begin{equation}
\begin{gathered}
\label{2.5}\beta_{n}^{(1)}=(-1)^n\,\left(3\left[\frac{n+1}{2}\right]-
1+(-1)^n\right),\\
\beta_{n}^{(2)}=(-1)^{n+1}\,\left(3\left[\frac{n}{2}\right]+1+(-1)^
{n+1}\right)
\end{gathered}
\end{equation}
with $[x]$ denoting the integer part of $x$. The only remaining
rational solution is the trivial solution $w(z;0)=0$.
\end{theorem}
\begin{proof}
While proving this theorem we will miss the dependence of
$w(z;\beta)$ on $\beta$. Any meromorphic solution of the equation
\eqref{1.1} can be represented as the ratio of two entire functions,
which we are going to construct. With the help of \eqref{2.2} we
understand that the function
\begin{equation}
\begin{gathered}
\label{t.1}f(z)=\exp(-\int^{z}ds_2\int^{s_2}w^2(s_1)ds_1)
\end{gathered}
\end{equation}
is entire and has a zero of multiplicity $e^2$ at $z=z_0$ whenever
$w(z)$ has a pole with residue $e$ at the same point $(e=$ $1$, $4$,
$-2$, $-3)$. The path of integration in \eqref{t.1} avoids the poles
of $w(z)$. Further the function $g(z)=w(z)f(z)$ is also entire.
Choosing in such a way entire functions we obtain the system of
equations satisfied by $f(z)$ and $g(z)$
\begin{equation}
\begin{gathered}
\label{t.2}ff_{zz}-f_z^2+g^2=0,\hfill \\
f^4g_{zzzz}-4f_zf^3g_{zzz}+(3g^2f^2+5g_zf^3+6f_z^2f^2
-5gf_zf^2)g_{zz} +(5gf^2 \hfill \\
-10f_zf^2)g_z^2+ (15fgf_z^2-4f_z^3f+5fg^3-16g^2f_zf)g_z+g^5
-5f_zg^4\hfill \\
+8ff_z^2g^3 -5f_z^3g^2+(f_z^4-zf^4)g-\beta f^5=0. \hfill
\end{gathered}
\end{equation}
In the case of a rational solution it can be set
\begin{equation}
\begin{gathered}
\label{t.3}f(z)=F(z)\exp(h(z)),\quad g(z)=G(z)\exp(h(z)),
\end{gathered}
\end{equation}
where $F(z)$, $G(z)$ are polynomials and $h(z)$ is an entire
function. Moreover it is straightforward to show that $h_{zz}=0$.
Substituting \eqref{t.3} into \eqref{t.2} yields a pair of equations
similar to \eqref{t.2} with $f(z)$ replaced by $F(z)$ and $g(z)$
replaced by $G(z)$. As far as a constant term is absent in
\eqref{2.2}, we see that
\begin{equation}
\begin{gathered}
\label{t.4}w^2(z)=\prod_{i=1}^{4}\prod_{k_i=1}^{l_i}\frac{i^2}{(z-z_{k_i})^2},
\end{gathered}
\end{equation}
where $w(z)$ is a rational solution having $l_i$ poles with residue
$i$ $(i=1,4)$ and $l_j$ poles with residue $-j$ $(j=2,3)$. Thus we
easily obtain
\begin{equation}
\begin{gathered}
\label{t.4a}F(z)=\prod_{i=1}^{4}\prod_{k_i=1}^{l_i}(z-z_{k_i})^{i^2}
\end{gathered}
\end{equation}
and
\begin{equation}
\begin{gathered}
\label{t.5}p\stackrel{def}{=}\deg(F(z))=\sum_{i=1,4}i^2l_i+\sum_{j=2,3}j^2l_j,\quad
\deg(G(z))=\deg(F(z))-1.
\end{gathered}
\end{equation}
The second correlation in \eqref{t.5} follows from \eqref{2.3}. The
polynomials $F(z)$, $G(z)$ can be written as
\begin{equation}
\begin{gathered}
\label{t.6}F(z)=\sum_{k=0}^{p}r_kz^{p-k},\quad
G(z)=\sum_{k=0}^{p-1}q_kz^{p-1-k}.
\end{gathered}
\end{equation}
Substituting \eqref{t.6} into \eqref{t.2} we find $q_0^2=p$,
$q_0=-\beta$. Consequently,
\begin{equation}
\begin{gathered}
\label{t.7}\sum_{i=1,4}i^2l_i+\sum_{j=2,3}j^2l_j=\beta^2.
\end{gathered}
\end{equation}
Further using the formula for the total sum of the residues of a
meromorphic function in the complex plane we get
\begin{equation}
\begin{gathered}
\label{t.8}\sum_{i=1,4}i\,l_i-\sum_{j=2,3}j\,l_j=-\beta.
\end{gathered}
\end{equation}
The correlations \eqref{t.7} and \eqref{t.6} should hold for any
rational solution. These correlations are not satisfied at
$\beta=1$. As a result the equation \eqref{1.1} does not have
rational solutions at $\beta=1$ and consequently at $\beta=1\pm3k$,
$k \in \mathbb{N}$. Assuming the contrary and applying the
B\"{a}cklund transformations \eqref{2.1} to a supposed rational
solution sufficient amount of times one can obtain the rational
solution at $\beta=1$. What is impossible. Thus the rational
solutions of \eqref{1.1} necessarily exist at $\beta \in
\mathbb{Z}/\{1\pm3k$, $k \in \mathbb{N}\cup0\}$, what follows from
the preceding remarks and from the correlation \eqref{t.8}, where
the left-hand side is an integer. The sufficiency follows from the
B\"{a}cklund transformations for \eqref{1.1}. Sequently applying the
B\"{a}cklund transformations $T_{2-\beta}$ and $T_{-1-\beta}$ to the
"seed" solution $w(z;0)=0$ we can construct the rational solutions
of \eqref{1.1} given by \eqref{2.4}. Starting with $T_{2-\beta}$ we
obtain the rational solutions at $\beta=\beta_n^{(1)}$, which we
will refer to as the first family. While starting with
$T_{-1-\beta}$ we obtain the rational solutions at
$\beta=\beta_n^{(2)}$, which we will refer to as the second family.
Note that $\{\beta_n^{(1)}\}\cup\{\beta_n^{(2)}\}\cup\{0\}$
$=\mathbb{Z}/\{1\pm3k$, $k \in \mathbb{N}\cup0\}$.
\end{proof}
\textit{Remark 1.} Expressions \eqref{2.5} can be rewritten as
\begin{equation}
\begin{gathered}
\label{2.7}\beta_{n}^{(1)}=(-1)^n\frac{3n}{2}+\frac{\delta_{n,odd}}{2},\hfill\\
\beta_{n}^{(2)}=(-1)^{n-1}\frac{3n}2+\frac{\delta_{n,odd}}{2}
\end{gathered}
\end{equation}
with $\delta_{n,odd}$ being the Kronecker delta
\begin{equation}\label{2.8}
\delta_{n,odd}=\left\{
\begin{gathered}
1,\quad n \,\,(mod 2)=1;\\
0,\quad n\,\,(mod 2)=0.
\end{gathered}
\right.
\end{equation}

Thus we see that the rational solutions of \eqref{1.1} can be
described with the help of two families of polynomials. The
polynomials $\{Q_n(z)\}$ we will call the first family and
$\{R_n(z)\}$ the second. By $p_n^{(1)}$ $(p_n^{(2)})$ denote the
degree of $Q_n(z)$ $(R_n(z))$. Analyzing the expression \eqref{2.4}
we understand that $Q_n(z)$ and $R_n(z)$ can be defined as monic
polynomials. Thus each polynomial can be presented in the form
\begin{equation}\begin{gathered}
\label{2.9}Q_n(z)=\sum_{k=0}^{p_{n}^{(1)}}A^{(1)}_{n,k}\,z^{p_{n}^{(1)}\,-\,k},\qquad
A^{(1)}_{n,0}=1,\\
R_n(z)=\sum_{k=0}^{p_{n}^{(2)}}A^{(2)}_{n,k}\,z^{p_{n}^{(2)}\,-\,k},\qquad
A^{(2)}_{n,0}=1.
\end{gathered}
\end{equation}
The first non-trivial solutions of \eqref{1.1} are $w(z;-1)=1/z$ and
$w(z;2)=-2/z$. Hence it can be set $Q_0(z)=R_0(z)=1$, $Q_1(z)=z$,
$R_1(z)=z^2$. Suppose $a^{(j)}_{n,k}$ $(1\leq k\leq p_{n}^{(j)})$
are the roots of the polynomial $Q_n(z)$ and $R_n(z)$, accordingly,
then by $s^{(j)}_{n,k}$ we denote the symmetric functions of the
roots
\begin{equation}
\label{2.10}s^{(j)}_{n,m}\stackrel{def}{=}\sum_{k=1}^{p_{n}^{(j)}}(a^{(j)}_{n,k})^m,\quad
m\geq 1,\quad j=1,2.
\end{equation}
Let us show that it is possible to derive the polynomials $Q_n(z)$,
$R_n(z)$ using the power expansion \eqref{2.3}. For convenience of
use let us present this series in the form
\begin{equation}
\label{2.11}w(z;\beta)=\sum_{m=0}^{\infty}c_{\beta,-(m+1)}z^{-m-1},
\end{equation}
where $c_{\beta,-(m+1)}=0$ $(m\geq1)$ unless $m$ is divisible by
$5$. Our next step is to express $s_{n,m}^{(j)}$ and $p_n^{(j)}$
through coefficients of the series \eqref{2.11}.

\begin{theorem}
\label{T:2.2} Let $c_{-m-1}^{(j)}(i)$ be the coefficient in
expansion \eqref{2.11} at  $\beta=\beta_i^{(j)}$. Then for each
$n\geq2$ and $j=1,2$ the following relations hold
\begin{equation}
\label{2.13a}p_n^{(j)}=(-1)^{j-1}\sum_{i=1}^{n}(-1)^{i-1}c_{-1}^{(j)}(i),
\end{equation}
\begin{equation}
\label{2.13}s_{n,m}^{(j)}=(-1)^{j-1}\sum_{i=2}^{n}(-1)^{i-1}c_{-(m+1)}^{(j)}(i),\quad
m\geq1.
\end{equation}
\end{theorem}

\begin{proof} Without loss of generality we will prove the theorem
for the polynomials $\{Q_n(z)\}$. Up to the end of the proof the
upper index $j$ will be omitted. As far as $Q_n(z)$ is a monic
polynomial, then it can be written in the form
\begin{equation}
\label{2.14}Q_n(z)=\prod_{k=1}^{p_{n}}(z-a_{n,k}).
\end{equation}
Note that possibly $a_{n,k}=a_{n,l}$, $k\neq l$. This equality
implies that
\begin{equation}
\label{2.15}\frac{Q_n^{'}(z)}{Q_n(z)}=\sum_{k=1}^{p_{n}}\frac1{z-a_{n,k}}.
\end{equation}
Substituting \eqref{2.15} into the expression \eqref{2.4} yields
\begin{equation}
\label{2.16}w(z;\beta_n)=(-1)^n\left(
\sum_{k=1}^{p_{n-1}}\frac1{z-a_{{n-1},k}}-\sum_{k=1}^{p_{n}}\frac1{z-a_{n,k}}\right).
\end{equation}
Expanding this function in a neighborhood of infinity we get
\begin{equation}
\begin{gathered}
\label{2.17}w(z;\beta_n)= (-1)^n
\left(\frac{b_{n-1}}{z}-\frac{b_n}{z}\right)+(-1)^n
\sum_{m=0}^{\infty}
\left[\sum_{k=1+b_{n-1}}^{p_{n-1}}(a_{n-1,k})^m\right.\\
\left.-\sum_{k=1+b_n }^{p_{n}}(a_{n,k})^m\right] z^{-(m+1)},\,\,
|z|>\max\{\tilde{a}_{n-1},\tilde{a}_{n} \},\,\,\\
\tilde{a}_{n}=\max\limits_{1\leq k \leq
p_{n}}\{|a_{n,k}|\},\,\,b_n=\sum_{k=1}^{p_{n}}\delta_{0,a_{n,k}},
\end{gathered}
\end{equation}
where $\delta_{0,a_{n,k}}$ is the Kronecker delta. The first or the
second term in \eqref{2.17} are present only if the polynomials
$Q_{n-1}(z)$, $Q_n(z)$ have zero roots, accordingly. In our
designations the previous expression can be rewritten as
\begin{equation}
\begin{gathered}
\label{2.18}w(z;\beta_n)=(-1)^{n-1}\frac{p_{n}-p_{n-1}}{z}+(-1)^{n-1}
\sum_{m=1}^{\infty}\left[s_{n,m}-
s_{n-1,m}\right]z^{-(m+1)},\\
|z|>\max\{\tilde{a}_{n-1},\tilde{a}_{n}\}.
\end{gathered}
\end{equation}
The absence of a zero term in sum is essential only at $m=0$.
Comparing expansions \eqref{2.18} and \eqref{2.11} we obtain the
equalities
\begin{equation}
\begin{gathered}
\label{2.19}
p_{n}-p_{n-1}=(-1)^{n-1}c_{-1}(n),\hfill\\
s_{n,m}-s_{n-1,m}=(-1)^{n-1}c_{-(m+1)}(n),\quad m\geq1.
\end{gathered}
\end{equation}
In these expressions $c_{-(m+1)}(n)\stackrel{def}{=}
c_{\beta_n,-(m+1)}$, $m\geq 0$. Decreasing the first index by one in
\eqref{2.19} and adding the result to the original one yields
\begin{equation}
\begin{gathered}
\label{2.20}p_{n}-p_{n-2}=(-1)^{n-1}[c_{-1}(n)-c_{-1}(n-1)],\hfill\\
s_{n,m}-s_{n-2,m}=(-1)^{n-1}[c_{-(m+1)}(n)-c_{-(m+1)}(n-1)].
\end{gathered}
\end{equation}
Note that $c_{\beta_1,-(m+1)}=0,\,m\geq1$ and $a_{1,1}=0$. Then
proceeding in such a way we get the required relations \eqref{2.13a}
and \eqref{2.13}.
\end{proof}

Since $c_{\beta,-1}=-\beta$, we get that the degrees of the
polynomials $Q_n(z)$ and $R_n(z)$ are
\begin{equation}\begin{gathered}
\label{2.21}p_{n}^{(1)}=\frac12\sum_{i=1}^{n}(3i-\delta_{i,odd})=
\frac{n(3n+2)-\delta_{n,odd}}{4},\hfill\\
p_{n}^{(2)}=\frac12\sum_{i=1}^{n}(3i+\delta_{i,odd})=\frac{n(3n+4)+\delta_{n,odd}}{4}.
\end{gathered}
\end{equation}
We also observe that expression \eqref{2.21} can be rewritten in
terms of $\beta_{n}^{(j)}$
\begin{equation}\begin{gathered}
\label{2.21a}p_{n}^{(1)}=\sum_{i=1}^{n}|\beta_i^{(1)}|=\frac{k(k+1)}{2}-\frac12\,
\left[\frac{k+1}{3}\right]-
\frac32\left[\frac{k+1}{3}\right]^2,\,k\stackrel{def}{=} |\beta_n^{(1)}|,\hfill\\
p_{n}^{(2)}=\sum_{i=1}^{n}|\beta_i^{(2)}|=\frac{k(k+1)}{2}+\frac12\,\left[\frac{k+2}
{3}\right]-
\frac32\left[\frac{k+2}{3}\right]^2,\,k\stackrel{def}{=}
|\beta_n^{(2)}|.
\end{gathered}
\end{equation}
The first few $\beta_{n}^{(j)}$ and $p_n^{(j)}$ are given in Table
\ref{t:1}.

\begin{table}[h]
    \center
    \caption{Values of $\beta_{n}^{(j)}$ and $p_n^{(j)}$ $(j=1,2)$. } \label{t:1}
    \begin{tabular}{|c|c|c|c|c|c|c|c|c|c|c|c|c|} 
        \hline
        $ n $& $1$ &$2$ &$3$ & $4$ & $5$ & $6$ & $7$ & $8$ & $9$ & $10$ & $11$ &
        $12$\\ \hline
        $ \beta_{n}^{(1)} $ & $-1$ &$3$ & $-4$ & $6$ & $-7$ & $9$ & $-10$ & $12$ & $-13$
        & $15$ & $-16$ &
        $18$\\ \hline
        $ p_{n}^{(1)} $ & $1$ &$4$ & $8$ & $14$ & $21$ & $30$ & $40$ & $52$ & $65$
        & $80$ & $96$ &
        $114$\\ \hline
         $ \beta_{n}^{(2)} $ & $2$ &$-3$ & $5$ & $-6$ & $8$ & $-9$ & $11$ & $-12$ &
         $14$ & $-15$ & $17$ &
        $-18$\\ \hline
        $ p_{n}^{(2)} $ & $2$ &$5$ & $10$ & $16$ & $24$ & $33$ & $44$ & $56$ & $70$ &
         $85$ & $102$ & $120$\\
        \hline
    \end{tabular}
\end{table}

Theorem \eqref{T:2.2} enables us to prove the following theorem.

\begin{theorem}
\label{T:1.1.} All the coefficients $A_{n,m}^{(1)}$
$(A_{n,m}^{(2)})$ of the polynomial $Q_n(z)$ $(R_n(z))$ can be
obtained with the help of $p_n^{(1)}$ $(p_n^{(2)})$ first
coefficients of the expansion \eqref{2.11} for the solutions of
\eqref{1.1}.
\end{theorem}

\begin{proof} Again we omit the upper index, when it does not cause
any contradiction. For every polynomial there exists a connection
between its coefficients and the symmetric functions of its roots
$s_{n,m}$. This connection is the following
\begin{equation}
\label{2.22}mA_{n,m}+s_{n,1}A_{n,m-1}+\ldots +s_{n,m}A_{n,0}=0,\quad
1\leq m\leq p_n.
\end{equation}
Taking into account that in our case $A_{n,0}=1$ we get
\begin{equation}
\label{2.23}A_{n,m}=-\frac{s_{n,m}+s_{n,m-1}A_{n,1}+\ldots
+s_{n,1}A_{n,m-1}}{m},\quad 1\leq m\leq p_n.
\end{equation}
The function $s_{n,m}$ can be derived using the expression
\eqref{2.13}. Hence recalling the fact that \eqref{2.11} is exactly
\eqref{2.4} we obtain
\begin{equation}
\begin{gathered}
\label{2.24}s_{n,m}=0,\quad m\in \textbf{N}\,/\,\{5l,\quad l\in
\textbf{N}\},\hfill\\
s_{n,5\,l}^{(j)}=(-1)^{j-1}\sum_{i=2}^{n}(-1)^{i-1}c_{-(5l+1)}^{(j)}(i),\quad
l\in \textbf{N}.
\end{gathered}
\end{equation}
Substituting this into \eqref{2.23} yields
\begin{equation}
\begin{gathered}
\label{2.25}A_{n,m}=0,\quad m\in\{1,2,\ldots
,n(n+1)/2\}\,/\,\{5l,\quad l\in \textbf{N}\},\hfill \\
A_{n,5\,l}=-\frac{1}{5l}\{s_{n,5\,l}+s_{n,5\,l-5}A_{n,5}+ \ldots
+s_{n,5}A_{n,5\,l-5}\},\, l\in \textbf{N},\,5l\leq p_n.
\end{gathered}
\end{equation}
Thus we see that the coefficients $A_{n,k}$ of the polynomial
$Q_n(z)$ $(R_n(z))$ are uniquely defined by coefficients
$c_{\beta_n,-{5l}-1}$ of the expansion \eqref{2.3}. This completes
the proof.
\end{proof}
\textit{Remark 1.} At given $n\geq2$ the functions $s_{n,m}^{(j)}$
$(m>p_{n}^{(j)})$ do not contain any new information about the roots
of corresponding polynomial $(Q_n(z)$ or $R_n(z))$ as the following
correlation holds
\begin{equation}
\label{2.25a}s_{n,m}^{(j)}+s_{n,m-1}^{(j)}A_{n,1}^{(j)}+\ldots
+s_{n,m-p_n}^{(j)}A_{n,p_n}^{(j)}=0,\, m>p_{n},\, m\in
\textbf{N},\,j=1,2.
\end{equation}

\textit{Remark 2.} Expression \eqref{2.25} defines the structure of
the polynomial $Q_n(z)$ $(R_n(z))$. Namely if $p_n^{(j)}$ is
divisible by $5$, then $Q_n(z)$ $(R_n(z))$ is a polynomial in $z^5$.
Otherwise (if $p_n^{(j)}$ is not divisible by $5$) $Q_n(z)/z^r$
$(R_n(z)/z^r)$ is a polynomial in $z^5$, where
$r=p_n^{(j)}\,(mod\,5)$.

\begin{table}[h]
    \center
    \caption{Polynomials $Q_n(z)$} \label{t:k40}
    \begin{tabular}{l} 
        \hline
        $ Q_0=1 $, \\
        $ Q_1=z $, \\
        $ Q_2=z^4$, \\
        $ Q_3=z^8 $, \\
        $ Q_4=z^4\,(z^5-504)^2 $, \\
        $ Q_5 = z\, ( {z}^{20}-3276\,{z}^{15}+6604416\,{z}^{10} +3328625664 \,{z}^{5}-
        119830523904 ) $, \\
        $ Q_{{6}}= ({z}^{15}-6552\,{z}^{10}-13208832\,{z}^{5}-951035904)^{2} $, \\
        $ Q_{{7}}={z}^{40}-29952\,{z}^{35}+203793408\,{z}^{30}+3066139754496\,{z}^{25}+ $ \\
        \qquad $ +5234197284126720\,{z}^{20}+36006491762989203456\,{z}^{15}- $ \\
        \qquad $ -3574462636834928197632\,{z}^{10}-7206116675859215246426112\,{z}^{5}+ $ \\
        \qquad $ + 129710100165465874435670016 $ \\
        \hline
    \end{tabular}
\end{table}

Calculating the coefficients $c_{\beta,-5l-1}$ in \eqref{2.3} and
the symmetric functions $s_{n,5l}^{(j)}$ one can find the
coefficients $A_{n,5l}^{(j)}$ of the polynomials $Q_n(z)$ and
$R_n(z)$. It can be proved that $c_{\beta,-5l-1}$ is a polynomial in
$\beta$. Hence, taking into account \eqref{2.7} we see that all the
coefficients $A_{n,5l}^{(j)}$ can be presented in the form
\begin{equation}
\begin{gathered}
\label{2.26}A_{n,5l}^{(j)}=B_{n,5l}^{(j)}+\delta_{n,odd}C_{n,5l}^{(j)}
\end{gathered}
\end{equation}
with $B_{n,5l}^{(j)}$ and $C_{n,5l}^{(j)}$ being polynomials in $n$.
For example,
\begin{equation}
\begin{gathered}
\label{2.27}A_{n,5}^{(1)}=-{\frac {3}{320}}\,n \left( 3\,n+8 \right)
\left( 3\,n+2 \right)
 \left( 3\,n-4 \right)  \left( {n}^{2}-4 \right) +\frac{\delta_{n,\,odd}}{320}\\
\times\,
 \left( {405}\,{n}^{4}+540\,{n}^{3}-495\,{n}^{2}-450\,n+495 \right)
\end{gathered}
\end{equation}
\begin{equation}
\begin{gathered}
\label{2.28}A_{n,5}^{(2)}=-{\frac {3}{320}}\,n \left( 3\,n+4 \right)
\left( 3\,n-2 \right)
 \left( 3\,n-8 \right)  \left( n+4 \right)  \left( n+2 \right) -
 \frac{\delta_{n,\,odd}}{320}\\
\times\, \left( 405\,{n}^{4}+1080\,{n}^{3} +315\,{n}^{2}-540\,n+315
 \right)
\end{gathered}
\end{equation}

\begin{table}[h]
    \center
    \caption{Polynomials $R_n(z)$} \label{t:k41}
    \begin{tabular}{l}
        \hline
        $ R_0=1 $, \\
        $ R_1=z^2 $, \\
        $ R_2=z^5+36 $, \\
        $ R_3=(z^5-144)^2 $, \\
        $ R_{{4}}=z \left( {z}^{15}-1152\,{z}^{10}+1824768\,{z}^{5}+131383296 \right) $, \\
        $ R_{{5}}={z}^{4} \left( {z}^{10}-3168\,{z}^{5}-3193344 \right) ^{2} $, \\
        $ R_{{6}}={z}^{8} \left({z}^{25}-15840\,{z}^{20}+63866880\,{z}^{15}+
        708155965440\,{z}^{10}+ \right. $ \\
        \qquad $ \left.+1922177762806726656 \right) $, \\
        $ R_{{7}}={z}^{4} ({z}^{20}-22176\,{z}^{15}-95001984\,{z}^{10}-
        902898855936\,{z}^{5}+
         $ \\
        \qquad $ +303374015594496) ^{2} $ \\
                \hline
    \end{tabular}
\end{table}
Additionally the coefficients $A_{2m-1,5l}^{(j)}$ and
$A_{2m,5l}^{(j)}$ are polynomials in $m$. Some of them for the first
family of polynomials are the following
\begin{equation}
\begin{gathered}
\label{2.29}A_{2m-1,5}^{(1)}=-\frac35\,m \left( m-1 \right)  \left(
m-2 \right)  \left( 3\,m+4 \right)
 \left( 3\,m+1 \right)  \left( 3\,m-2 \right)
\end{gathered}
\end{equation}
\begin{equation}
\begin{gathered}
\label{2.30}A_{2m,5}^{(1)}=-\frac35\,m  \left( {m}^{2}-1
\right)\left( 3\,m+4 \right)  \left( 3\,m+1 \right)  \left( 3\,m-2
 \right)
\end{gathered}
\end{equation}
\begin{equation}
\begin{gathered}
\label{2.31}A_{2m-1,10}^{(1)}={\frac {9}{50}}\,m \left( m^2-1
\right) \left( m-2 \right)  \left( 3\,m +4 \right)  \left( 3\,m+1
\right) \left( 3\,m-2 \right) \\
\times \left( 3\,m- 5 \right) \left(
9\,{m}^{4}-12\,{m}^{3}-143\,{m}^ {2}+98\,m+1008 \right)
\end{gathered}
\end{equation}
\begin{equation}
\begin{gathered}
\label{2.32}A_{2m,10}^{(1)}={\frac {9}{50}}\,m  \left( {m}^{2}-1
 \right)\left( 3\,m+4 \right)
\left( 3\,m+1 \right)
 \left( 3\,m-2 \right)
  \left( 3\,{m}^{2}+m-9 \right) \\
\times   \left( 9\,{m}^{ 4}+6\,{m}^{3}-125\,{m}^{2}-42\,m+560
\right)
\end{gathered}
\end{equation}
The same coefficients for the second family are given by
\begin{equation}
\begin{gathered}
\label{2.33}A_{2m-1,5}^{(2)}=-\frac35\,m \left( m^2-1 \right) \left(
3\,m+2 \right)  \left( 3\,m-1
 \right)  \left( 3\,m-4 \right)
\end{gathered}
\end{equation}
\begin{equation}
\begin{gathered}
\label{2.34}A_{2m,5}^{(2)}=-\frac35\,m \left( 3\,m+2 \right)  \left(
3\,m-1 \right)  \left( 3\,m-4
 \right)  \left( m+2 \right)  \left( m+1 \right)
\end{gathered}
\end{equation}
\begin{equation}
\begin{gathered}
\label{2.35}A_{2m-1,10}^{(2)}={\frac {9}{50}}\,m \left( m^2-1
\right) \left( 3\,m+2 \right)  \left( 3 \,m-1 \right)  \left( 3\,m-4
\right) \left( 3\,{m }^{2}-m-9 \right)\\
  \times\left(
9\,{m}^{4}-6\,{m}^{3}-125\,{m}^{2}+42\,m+560
 \right)
\end{gathered}
\end{equation}
\begin{equation}
\begin{gathered}
\label{2.36}A_{2m,10}^{(2)}={\frac {9}{50}}\,m \left( m^2-1 \right)
\left( 3\,m+5 \right)  \left( 3 \,m+2 \right)  \left( m+2 \right)
\left( 3\,m-1 \right)
  \left( 3\,m-4 \right)\\
    \times\left( 9\,{m}^{4}+12\,{m}^{3}-143\,{m
}^{2}-98\,m+1008 \right)
\end{gathered}
\end{equation}
Note that the expressions \eqref{2.29}, \eqref{2.30} are equivalent
to \eqref{2.27}. Similarly, \eqref{2.33}, \eqref{2.34} are
equivalent to \eqref{2.28}. The polynomials $Q_n(z)$ and $R_n(z)$
can be written in the form
\begin{equation}
\begin{gathered}
\label{2.37}Q_n(z)=\sum_{l=0}^{[p_n^{(1)}/5]}A_{n,5l}^{(1)}z^{p_n^{(1)}-5l},
\quad A_{n,0}^{(1)}=1,\\
R_n(z)=\sum_{l=0}^{[p_n^{(2)}/5]}A_{n,5l}^{(2)}z^{p_n^{(2)}-5l},\quad
A_{n,0}^{(2)}=1.
\end{gathered}
\end{equation}

The first few of them are gathered in Tables \ref{t:k40} and
\ref{t:k41}.

\section{Rational solutions of the equation studied}

Substituting the polynomials $Q_{n-1}(z)$, $Q_{n}(z)$ and
$R_{n-1}(z)$, $R_{n}(z)$ into \eqref{2.5} we obtain the rational
solutions of the equation \eqref{1.1}. Some of them are given below.
\begin{equation}
\begin{gathered}
\label{k43}w(z,-1)=\frac{1}{z},\quad
w(z;3)=-\frac{3}{z},\quad\,w(z,-4)=\frac{4}{z},
\end{gathered}
\end{equation}

\begin{equation}
\begin{gathered}
\label{k44}w \left( z,6 \right) =-\,{\frac {6\,({z}^{5}+336)}{z
\left( {z}^{5}-504 \right) }},
\end{gathered}
\end{equation}

\begin{equation}
\begin{gathered}
\label{k45}w(z,2)=-\frac{2}{z},\quad w \left( z,-3 \right) =\,{\frac
{3({z}^{5}-24)}{z \left( {z}^{5}+36
 \right) }}
\end{gathered}
\end{equation}

\begin{equation}
\begin{gathered}
\label{k46}w \left( z,5 \right) =-\,{\frac {5\,{z}^{4} \left(
{z}^{5}+216 \right) } { \left( {z}^{5}+36 \right)  \left(
{z}^{5}-144 \right) }},
\end{gathered}
\end{equation}

\begin{equation}
\begin{gathered}
\label{k47}w \left( z,-6 \right) ={\frac
{6({z}^{20}-576{z}^{15}-912384{z}^{10}-459841536{z}^{5}-3153199104)}{z
\left( {z}^{15}-1152\,{z}^{10}+ 1824768\,{z}^{5}+131383296 \right)
\left( {z}^{5}-144 \right) }}
\end{gathered}
\end{equation}

Solutions \eqref{k43}, \eqref{k44} belong to the first family, while
\eqref{k45} -- \eqref{k47} to the second.

\section {Conclusion}

In this paper we have studied rational solutions of the equation
\eqref{1.1}. We have given necessary and sufficient condition for
the existence of the rational solutions. Further we have derived
special polynomials associated with these solutions. This property
of the equation \eqref{1.1} is similar to that of $P_2-P_6$. The
rational solutions have been subdivided into two sequences, each
with its own family of polynomials. Using the power expansion in a
neighborhood of infinity for the solutions of \eqref{1.1} we have
found the degrees of the polynomials and formulas for their
coefficients. We have shown that special solutions of the Fordy -
Gibbons, the Caudrey - Dodd - Gibbon and the Kaup - Kupershmidt
equations can be expressed via solutions of \eqref{1.1}.
Consequently, there exist solutions of these equations in terms of
the polynomials constructed.

\section {Acknowledgments}

This work was supported by the International Science and Technology
Center under Project B 1213.

\end{document}